# Log Management Support for Recovery in Mobile Computing Environment

*J.C. Miraclin Joyce Pamila,*
*Senior Lecturer, CSE & IT Dept,*
*Government College of Technology,*
*Coimbatore, India.*
miraclin_2000@yahoo.com

*K. Thanushkodi,*
*Principal,*
*Akshaya College of Engineering and Technology,*
*Coimbatore, India.*
thanush_dr@rediffmail.com

*Abstract* **-** **Rapid and innovative improvement in wireless communication technologies has led to an increase in the demand for mobile internet transactions. However, internet access from mobile devices is very expensive due to limited bandwidth available on wireless links and high mobility rate of mobile hosts. When a user executes a transaction with a web portal from a mobile device, the disconnection necessitates failure of the transaction or redoing all the steps after reconnection, to get back into consistent application state. Thus considering challenges in wireless mobile networks, a new log management scheme is proposed for recovery of mobile transactions.**

**In this proposed approach, the model parameters that affect application state recovery are analyzed. The proposed scheme is compared with the existing Lazy and Pessimistic scheme and a trade off analysis between the cost invested to manage log and the return of investment in terms of improved failure recoverability is made. From the analysis, the best checkpoint interval period that yields the best return of investment is identified.**

*Keywords – log, recovery, mobile environment.*

## I. INTRODUCTION

Improvement in quality, security and reliability of cellular services, facilitates internet access from the mobile devices. A mobile host (MH) engaged in a client-server application, may easily fail because of limited network resources. Due to its potential applicability, failure recovery of client-server applications needs considerable attention.

Checkpoint and message logging protocols are designed for saving the execution state of a mobile application, so that when a MH recovers from a failure, the mobile application can roll back to the last saved consistent state, and restart execution with recovery guarantees. The existing protocols assume that the MH's disk storage is not stable and thus checkpoint and log information are stored at the base stations [1], [6].

Two broad categories of mobile checkpoint protocols i.e. coordinated and uncoordinated have been proposed in the literature. Coordinated protocols, are suitable for MHs that run distributed application, and MH must coordinate their local checkpoints to ensure a consistent and recoverable global checkpoint [7]. Uncoordinated protocols are more applicable to mobile applications involving only a single client MH, and the MH can independently checkpoint its local state.

Pradhan, Krishna, and Vaidya [3] proposed two uncoordinated checkpoint protocols: No-logging and Logging approaches. The No-logging approach requires the MH to create a new checkpoint every time it has write-event that modifies the state of the application. The Logging approach, creates checkpoints only periodically, and logs all write-events which occur in between two checkpoints. When a MH recovers from a failure, it will retrieve the checkpoint along with log entries saved, to start the recovery process. Performance analysis of Logging versus No-logging was reported in [3].

## II. RELATED WORK

Global checkpoint based schemes [1] consider distributed applications running on multiple mobile hosts. Hence asynchronous recovery schemes [5], [6] are best suited than the schemes of [12] which require synchronization messages between participating processes.

Lazy and Pessimistic schemes are reported in [8]. In a lazy scheme, logs are stored in the Base station (BS) and if mobile host moves to a new BS, a pointer to the old BS gets stored in the new BS. Pointers can be used to recover the log distributed over several BS's. This scheme has the advantage that it incurs relatively less network overhead during handoff as no log information needs to be transferred. But this scheme has a large recovery time. In the pessimistic scheme, the entire log and checkpoint record, if any, are transferred at each handoff. Hence, the recovery is fast but each handoff requires large volume of data transfer.

The work reported in [2], [9] present schemes based on the mobile host's movement using independent check pointing and pessimistic logging. In the distance-based scheme, the contents that are distributed are unified when the distance covered by mobile host increases above a predefined value. After unifying the log, the distance or handoff counter is reset. These schemes are a trade off between lazy and the pessimistic strategies. The schemes discussed so far do not consider the case where a mobile host recovers in a base station different than the one in which it has crashed.

The mobile agent based framework proposed in [4] addresses this problem. This facilitates seamless logging of application activities for recovery from transactions failure.

All the previous works discuss about storing log in Base Station [4]. In the proposed approach Base station Controller is selected for storing the log information rather than Base Station.

### III. REFERENCE ARCHITECTURE

Fig. 1 illustrates the reference architecture. It is a client/server architecture system based on GSM [13]. A mobile network contains, a fixed backbone network and wireless network. A host that can move while retaining its network connection is a mobile host. A static network consists of fixed hosts and base stations (BS),

BS that interacts with MH and with wired network acts as a gateway between wired and wireless networks. Two or more base stations are controlled by a Base Station Controller (BSC). Similarly Mobile Switching Center (MSC) will control two or more BSC's.

BS comprises all radio equipment needed for radio transmission and forms a radio cell. It is connected to MH via radio link and to BSC via a high speed wired link. BS will act as a switch to which a mobile Host can communicate within a limited geographical region referred as a cell. Due to mobility, the MH may cross the boundary between two cells while being active and the process is known as handoff.

BSC manages the Base Stations. It reserves radio frequencies, handles the handover from one BS to another within the BSC region, and also performs paging of the MH. MSC is high performance digital ISDN switch and is equipped with Home Location Register and Visitor Location Register for storing the location information of the mobile hosts.

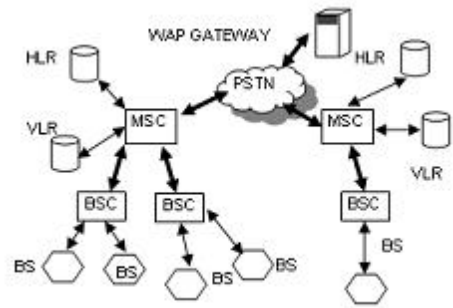

Fig 1. Reference Architecture

As wireless bandwidth is constrained in cellular networks, HTTP is not feasible for MH to access internet.Therefore, WAP-enabled devices (MH) communicate via a WAP gateway [14].

The gateway turns the requests into standard web-based requests according to WAP specifications. The gateway acting as an internet client sends the request to a server providing WAP content.

### IV. RECOVERY MECHANISM

There are several factors that affect the recovery [1].

A. *Failure Rate of the Host*

The failure of MH due to weak wireless link or less battery power etc., are purely random in nature. If failures are more, the transaction has to roll back every time when MH recovers from failure and thus total execution period of transaction gets increased. Generally MH failure rate is approximated with exponential distribution or Poisson distribution.

B. *Log Size*

Transmission of data consumes twice as much power as receiving the same amount of data. So only essential write events are to be logged to reduce size of the log.

C. *Memory Constraints*

Storing the log of each MH at the BSC might use up a lot of memory space on the BSC. It is necessary to evaluate average memory requirements based on log size and the recovery schemes used.

D. *Recovery Time*

The time required to recover a process upon failure depends on the recovery scheme and method used for logging the write events.

*E. Log retrieval Cost*

The cost invested to retrieve the log information upon failure of a transaction depends on the amount of log distribution. If the log is distributed in more places, the retrieval cost and the recovery cost increase.

## V. PROPOSED LOG MANAGEMENT SCHEME

Here the log information is stored in BSC. The area covered by a single BSC is referred as a REGION. It is assumed that a Tracking agent present in BSC will query the HLR or VLR for the location update of mobile host in which transaction is initiated. By using this agent the problem of recovery of a mobile host in a BS different than the one in which it crashed is addressed.

*A. Intra BSC Management*

When MH is moving from one BS to other BS which is connected to same BSC, no log information is transferred as the log is in BSC. Therefore the handoff cost is reduced drastically.

*B. Inter BSC Management*

Every MH carries following information for the purpose of registration.
1. Previous BSC identity (PBSCid)
2. Own identity (MHid)

When a MH registers with a message Connect (MHid, PBSCid) to new BSC which is not the Home Base Station Controller (HBSC), then the new BSC informs HBSC about its reachability, by sending message containing MHid and HBSC and its own identity BSCid.

Now since this message is received by HBSC, it will transfer the entire log present in it to the current BSC.

*C. Log Transfer from Mobile cache to BSC*

MH transfers the entire log to the BSC as follows:
1. If MH cache is exhausted, immediately entire log will be copied to current BSC.
2. When Mobile Host moves away from the current BSC and system detects handoff, the MH will copy the entire log to BSC and the log would be appended to the previous log file.

## VI. MODELLING AND METRICS

In this section mathematical equations for different performance metrics are analyzed [8].

*A. Handoff Modeling*

The interval between two handoffs is referred to as handoff interval. A handoff interval can be represented using a 3 state discrete markov chain [8] as presented below.

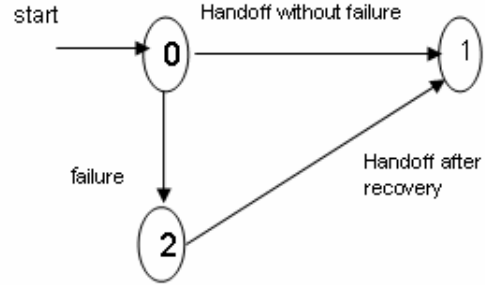

Fig. 2 Markov Chain Representation

In Fig. 2 State 0 is the initial state when the handoff interval begins. During the handoff interval, the host receives messages. Depending upon the state-saving scheme, the host either takes a checkpoint or logs the write events.

A transition from state 0 to state 1 occurs if the handoff interval is completed without failure. If a failure occurs during the handoff interval, a transition is made from state 0 to state 2. After state 2 is entered, a transition occurs to state 1 once the handoff interval is completed.

To simplify the analysis, it is assumed that, at most, one failure occurs during a handoff interval. This assumption does not significantly affect the results when the average handoff interval is small, compared to the mean time to failure.

*B. Terms and Notations*

- ❖ $\lambda$ - Log arrival rate
- ❖ $\mu$ - Handoff rate
- ❖ r - Ratio of the transfer time in the wired network to the transfer time in the wireless network.
- ❖ $\eta$ - Average log size
- ❖ $T_c$ – checkpoint interval.
- ❖ k - Number of write events per checkpoint
- ❖ $N_c$ – Number of checkpoints in t time units.
- ❖ $N_l$ - number of messages logged in time t.
- ❖ $C_c$ – Average transfer cost of a checkpoint state over one hop of the wired network.
- ❖ $C_l$ - Average transfer cost of an application message over one hop of the wired network.

- $C_m$ – Average transfer cost a control message over one hop of the wired network.
- T- Time to load last checkpoint
- $T_1$ – Time to load last log information
- α – wireless link cost
- ρ – wired link cost

Probability for a handoff without failure
$$P_{01} = 1 - (\lambda / (\lambda + \mu)). \quad (1)$$
Probability of failure within
$$\text{handoff } P_{02} = \lambda / (\lambda + \mu) \quad (2)$$

*C. Performance metrics*

*1) Handoff Cost:* In the proposed scheme, the message log will contain the write events that have been processed since the last checkpoint. For each logging operation, there is a cost for the acknowledgement message sent by the BSC to the MH. Thus handoff cost includes the cost of transferring the checkpoint state, message log, and an acknowledgement [8].
Average handoff cost is $C_h = (\eta C_l + C_C + C_m)$. (3)
Handoffs being Poisson process $\eta = (k-1)/2$. (4)
Thus the total Handoff cost [8] is
$$C_{01} = r\alpha C_c/k + \rho r\alpha C_l + \rho r\alpha C_m + \eta C_l + C_C + C_m \quad (5)$$

*2) Recovery Cost:* The recovery cost is the cost of transmitting a request message from the MH to the BSC, and the cost of transmitting the checkpoint and log over one hop of the wireless network.
For Poisson failure arrivals [8], $\eta = (k-1)/2$.
Therefore $C_r = r(\eta C_l + C_c + C_m)$ (6)

*3) Total Cost:* This is the expected cost incurred during a handoff interval with and without failure. The total cost is determined as follows
$C_t = P_{01} C_{01} + P_{02} C_r$ (7)

*D. Failure Recoverability*

In this section tradeoff involved between the cost invested for maintaining the checkpoint and log versus the resulting recovery probability gained when a failure occurs is analyzed.
*Failure Recoverability versus Cost Ratio* (*FRCR*) parameter [3] defined as the ratio of the difference in recovery probability to the difference in cost invested by these two strategies.
Then, $FRCR = (P_{prop} - P_{Lazy}) / (C_{prop} - C_{lazy})$. (8)
When given set of parameters values, $P_{prop}$ and $P_{lazy}$ are calculated. The cost invested by the two strategies is the cost incurred due to handoff. The proposed system will transfer all log and checkpoint information to the current BSC while Lazy method just establishes the link to previous base station.

The average number of checkpoints before failure is given by $(1/\lambda)/(T_c)$. The average number of moves crossing BSC's or BS's boundaries between two consecutive checkpoint is given by $T_c * \mu$. The total no of log entry transfer operations required by proposed system between two consecutive check points are given as

$$\sum_{n=1}^{Tc*\lambda}(1/\lambda)/(1/\mu)*n \quad (9)$$

The total cost invested by proposed system is $C_{prop}$
$$= (1/\lambda)/Tc * \{r*T*Tc*\lambda + r*T1*\sum_{n=1}^{tc*\lambda}(1/\lambda)/(1/\mu)*n \quad (10)$$
The cost invested by Lazy scheme[8] is
$$C_{lazy} = (1/\lambda)/Tc*(Tc*\lambda*Cp) \quad (11)$$

**VII. PERFORMANCE ANALYSIS**

The proposed scheme is implemented in Network Simulator 2 (NS2). In this section the proposed method is compared with the lazy and pessimistic methods.

*A. Comparison of handoff cost*
The handoff cost for lazy scheme is very low, as no log information is transferred during handoff. So irrespective of the Mobility rate, the handoff cost will remain same for Lazy scheme. The handoff cost for pessimistic is very high when compared with all schemes because for every handoff, the total log and check point information are to be transferred to the current BSC.

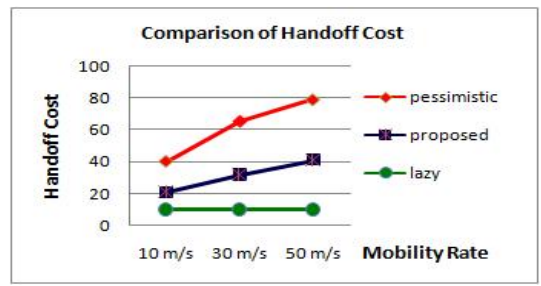

Fig. 3 Handoff cost of three strategies

The handoff cost for proposed scheme is low when compared to pessimistic, but high when compared to lazy scheme. But since the log and checkpoint information are not carried along with MH and after every checkpoint the log is purged the handoff cost for the proposed scheme is moderate.

Though mobility rate increases the handoff cost increment is very less when compared with the pessimistic scheme. The tradeoff between lazy, pessimistic and the proposed schemes are shown in Fig.3.

*B. Comparison of Recovery Cost*

From Fig. 4, since during recovery the logs are to be collected and this causes increase in recovery cost of lazy scheme. The increase in recovery cost will be more if the mobility rate increases. Recovery cost for pessimistic scheme is very low as the entire log information is present at the current base station.

Recovery cost for proposed scheme is also very low as the entire log information is present at the current BSC. But the recovery cost will be higher than the pessimistic scheme when recovery is in the same base station where mobile node got failed.

If recovery is in other Base station controller then the recovery cost for proposed is lower than the pessimistic scheme.

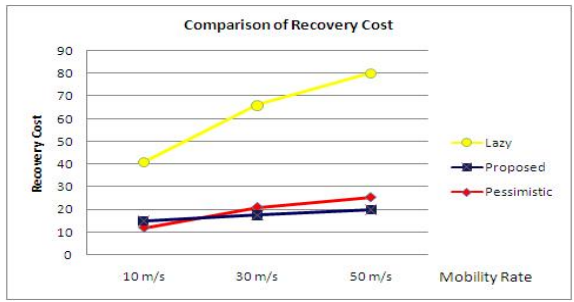

Fig. 4 Recovery cost of three strategies

*C. Comparison of Total Cost*

Total cost is sum of handoff cost and recovery cost. The total cost comparison shows that total cost incurred for the proposed scheme is comparatively very low.

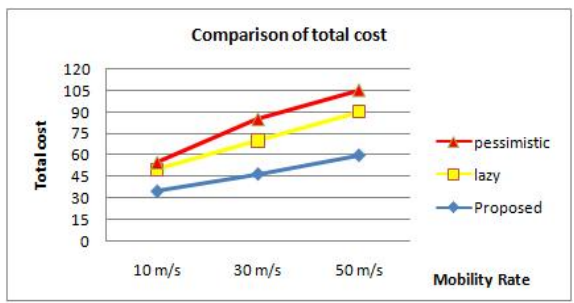

Fig. 5 Total Cost of three strategies

*D. Comparison of Recovery Probability*

Fig. 6 shows the effect of log arrival rate on failure recoverability. As observed, the system recovery probability decreases dramatically as the log arrival rate increases for all the schemes.

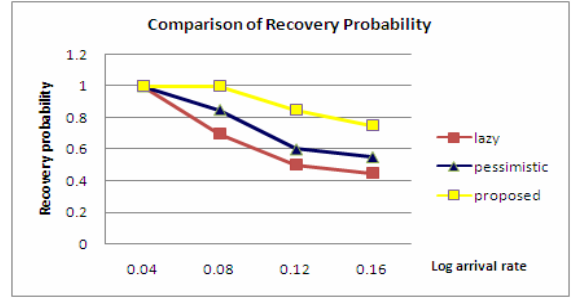

Fig. 6 Comparison of Recovery probability

For the proposed scheme also, the recovery probability decreases but decrease is less, compared with the other schemes. As the log information is stored at BSC, the probability is very high to recover in the same BSC. Even if it gets recovered in other BSC the entire log is present in the previous BSC. So the recovery probability is better for proposed scheme when compared with the other schemes.

*E. Failure Recoverability Vs Cost Analysis*

If the checkpoint interval is very short, all log entries since the last checkpoint as well as the last checkpoint itself are likely to reside in the current BSC, making the failure recoverability of both strategies virtually the same.

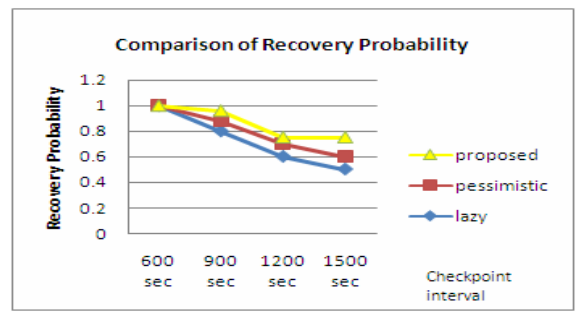

Fig. 7 Comparison of Recovery Probability

As the checkpoint interval increases, the number of log entries accumulated between two consecutive checkpoints becomes more substantial, thus resulting in an increase FRCR.

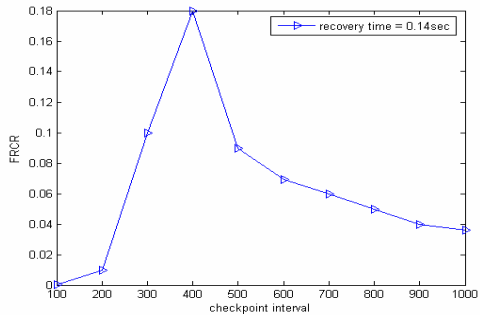

Fig. 8 Behavior of FRCR

When the checkpoint interval is very long, however, the improvement in failure recoverability cannot catch up with the increase in the cost investment difference, thus resulting in a decline in FRCR.

## VIII. CONCLUSION

The proposed log management scheme for mobile computing system reduces total cost for recovery from the failure when compared with the existing Lazy and pessimistic schemes.
The proposed technique also ensures recovery from different BS other than in which it has failed. The proposed scheme controls the handoff cost, log retrieval cost and failure recovery time. As a result of the analysis, the proposed scheme well suits when the mobility rate of the mobile host is very high.